\documentclass[%
 reprint,prd,
 amsmath,amssymb,longbibliography,aps,
]{revtex4-2}

\usepackage[margin=1.5cm]{geometry}
\usepackage{graphicx} 

\usepackage{subfigure}
\usepackage{dcolumn}
\usepackage{bm}
\usepackage{hyperref}
\usepackage{xcolor}
\usepackage{amsmath, amssymb, amsfonts}
\usepackage{mathtools}
\usepackage{mathrsfs}
\usepackage{comment}
\usepackage{amsthm}
\usepackage{ dsfont }
\usepackage{footnote}

\newcommand{\addappendix}{%
  \section*{\appendixname}
  \addcontentsline{toc}{section}{\appendixname}
  \counterwithin*{figure}{section}
  \stepcounter{section}
  \renewcommand{\thesection}{A}
  \renewcommand{\thefigure}{\thesection.\arabic{figure}}
}

\definecolor{purple1}{rgb}{128,0,128}

\newcommand{\nn}{\nonumber\\}
\newcommand{\bea}{\begin{eqnarray}}
\newcommand{\ea}{\end{eqnarray}}
\newcommand{\ord}{{\cal O}}

\theoremstyle{definition}

\definecolor{dfcol}{cmyk}{1, 0.2108, 0.13, 0.3}
\newcommand{\df}[1]{\ifthenelse{\boolean{}}{\textcolor{dfcol}{[{\bf DF}: #1]}}{}}

\renewcommand{\[}{\begin{equation}}
\renewcommand{\]}{\end{equation}}

\definecolor{mygray}{gray}{0.6}

\begin{document}
\title{{Quantum nonlinear effects 
in the number-conserving analogue gravity\\ of  Bose-Einstein condensates}}
\author{Kunal Pal}
\affiliation{%
Seoul National University, Department of Physics and Astronomy, Center for Theoretical Physics, Seoul 08826, Korea
}%
\author{Uwe R. Fischer}%
\affiliation{%
Seoul National University, Department of Physics and Astronomy, Center for Theoretical Physics, Seoul 08826, Korea
}

\date{\today}

\begin{abstract}
We consider the quantum dynamics of Bose-Einstein condensates at absolute zero,  and 
{demonstrate} that an analogue gravity model going beyond the standard linearized analogue gravity 
paradigm \`a la Unruh must take into account the backreaction of 
quasiparticle excitations onto the condensate background. This requires that one expands
to second order in perturbation amplitude and thus takes the intrinsic nonlinearity of the theory into account. 
It is shown that, as a result, significant modifications of the standard paradigm occur.
 In particular, to obtain a fully Lorentz-covariant equation in curved spacetime for second-order perturbations, we demonstrate that it is necessary to introduce, to leading order in powers of the formal mean-field 
expansion parameter $N^{-1/2}$ (where $N$ is total particle number), a quantum-fluctuation-renormalized spacetime metric which substantially differs from the Unruh acoustic spacetime metric and, to subleading order $1/N$, two emergent vector fields and a mass term. 
Both the renormalized metric as well as the vector fields and the mass then keep track 
of the backreaction of the quasiparticles onto the condensate up to the order in powers of $N^{-1/2}$ considered.  Finally, we apply our formalism to an analogue-cosmological Friedmann-Lema\^{i}tre-Robertson-Walker metric and establish its renormalized form due to the quantum many-body backreaction exerted by the excitation cloud. 
\end{abstract}

\maketitle

\section{Introduction}
The primary motivation behind the analogue gravity  programme is the experimental simulation of 
fundamental features of field quantization in curved spacetimes \cite{Birrell:1982ix} which are typically impossible to detect in real  gravitational fields \cite{Unruh1980,Visser:1997ux,Barcelo:2005fc}. Starting with the work of Unruh \cite{Unruh1980}, which put forth the idea that Hawking radiation can be observed in a lab fluid, analogue gravity models have now a rich history and found widespread applications to simulate various aspects of quantum field theory in curved spacetime cf., e.g., \cite{Jacobson1991,Unruh:1994je,Jacobson1996,Barcelo:2000tg, Barcelo:2003wu, Fischer:2004bf, Fedichev:2003id, Fedichev:2003bv,Macher,Kurita:2010bu,Bilic:1999sq}. A most recent review, 
containing numerous further references, is \cite{NatReview}; for a quick overview, a recent executive 
summary of analogue gravity's major agenda has been provided in Ref.~\cite{BarceloNature}.  

The analogue gravity idea of Unruh has led to the first observations of phenomena related to the paradigmatic quantum Hawking effect in the experimentally realizable physical system Bose-Einstein condensate (BEC), 
an ultracold quantum gas of interacting bosons \cite{Steinhauer16,Munoz,Kolobov2021}, the analogue gravity 
arena we focus on in the present study. Other BEC 
analogue gravity experiments have probed aspects of cosmology, such as Hubble
friction and preheating \cite{Eckel}, as well as phenomena related to cosmological particle production \cite{Hung,Viermann}.

One crucial aspect of probing quantized fields in an arbitrary curved spacetime that is often neglected is the possible effect of the backreaction of the fluctuating field on the background itself. The 
relative magnitude of backreaction, say in terms of the contribution to the semiclassical energy-momentum tensor, 
must be much smaller than that of the gravitating matter field for a clear separation between what constitutes  ``background" and ``fluctuation"  to be maintained. To the lowest order, as in the original prediction of the Hawking effect, backreaction is indeed conventionally neglected \cite{Hawking1974},  or at most a toy model for backreaction is proposed \cite{Hawking1975,Maia2007}. 
However,  a completely self-consistent treatment of any quantum field theory of a many-body system 
(or, similarly, of any candidate for an emergent unitary quantum field theory of gravity),   
should accommodate backreaction due to quantum-mechanical fluctuations 
created by its dynamical evolution.
In this regard, analogue gravity models based on BECs represent an ideal testbed to explore the possible role of quantum backreaction, as these systems afford a 
fully controlled description in terms of   
Bogoliubov theory \cite{Bogoliubov1947,Pethick2008bose,Pitaevskii2016bose}. 

It is of major importance in what follows that for a proper treatment of backreaction problems in quantum 
systems such as BECs, as was first pointed out in \cite{schutzhold2005quantum}, 
 {it is necessary to maintain 
particle-number conservation, cf.~Refs.~\cite{Girardeau1959theory,Gardiner:1997yk,castin1998low,gardiner2007number}, which are  
obtaining variants of number-conserving formulations of Bogoliubov theory 
which go beyond the symmetry-breaking traditional Bogoliubov approach.} 
In \cite{schutzhold2005quantum} the backreaction problem was defined by employing an expansion of the field operator in powers of $N^{1/2}$, where $N$ is the total number of particles (here
either atoms or molecules) in the system. Then the backreaction of the quantum fluctuations can be elegantly treated as the subleading correction to the leading order in $N^{1/2}$ Gross-Pitaevski\v\i\/     
equation (GPE) [which is equivalent to a nonrelativistic variant of $\phi^4$ theory],  
incorporated by the modified coupled dynamical equation of the condensate and the noncondensate part.  
A particular lesson to be drawn from the analysis of \cite{schutzhold2005quantum} is that  in a 
number-conserving 
description of the backreaction problem, the usual assumption of taking the expectation value of the quantized energy-momentum tensor, $\hat{T}_{\mu\nu}$, in a suitable quantum state, cf.~Ref.~\cite{BalbinotPRD}, 
does not necessarily reproduce the correct, that is {\em observable} physical backreaction force. 

{The consequences of using a number-conserving approach were further explored in \cite{baak2022number},} where the solution for backreaction as a Cauchy problem was formulated, and solved for a dynamical BEC configuration. Importantly, it was shown in \cite{baak2022number,Ribeiro:2024vem} that there can exist a nonzero flux of  condensate particles even in the absence of any particle flux due to quantum depletion, solely due to backreaction. 
 Backreaction for a solitonic solution of the nonlinear Schr\"odinger equation 
 was also studied most recently in \cite{Baak:2024ajn}.

{In the present study, our aim is to build an analogue gravity model based on the  
number-conserving approach of \cite{Baak:2024ajn}, which 
incorporates the coupled dynamics (particle exchange) of the condensate and the depleted cloud.} 
As we will explain in the sequel, the following is based on two key features that emerge from a consistent treatment of the backreaction of the depleted cloud {onto} the background  \cite{schutzhold2005quantum,Fischer2007,Schutzhold:2007fg,baak2022number}. First of all, due to number conservation, there is a nonvanishing particle flux from the condensate to the depletion part. 
As a result, 
the continuity equations for density and current of both condensate and depletion
contain drain (source) terms. 
In addition, the background to build an analogue gravity description now has a subleading contribution that represents a correction to the usual GPE. Consequently, as we argue, 
the standard approach of using strictly linear perturbation fields (see for example Visser's pedagogical paper \cite{Visser:1997ux}) is not consistent with the order (in the perturbation expansion in $N$) of the field operator 
we will work with, ${\mathcal O}(N^{-1/2})$, which will be demonstrated to be the minimal expansion order 
necessary for number conservation to be maintained. 
We are thus required to go beyond the linear perturbation paradigm of Unruh. 
On the classical level, the latter was developed in Ref.~\cite{datta2022analogue}, where it was shown how to extract an effective metric description from a generic nonlinear perturbation, at the cost of modifying the definition of background, also cf.~Ref.~\cite{fernandes2022dynamical}.

We will argue that the above procedure engenders a significant departure from the established formulation of the fluid perturbation equations as those of a scalar field in curved spacetime and that it is not much less straightforward to conclude that such a description can be envisaged if both backreaction and nonlinearity are taken into account. 
We will indeed demonstrate that for a Lorentz-invariant form of the nonlinear perturbation equation for a scalar field to exist, the latter will acquire, in subleading order, a mass term and two 
``emergent" vector fields. To leading order in 
the large $N$ expansion, the effective curved background metric acquires corrections stemming from 
the exchange of particles between condensate and excitation cloud. 
As our analysis in this paper is based on the Bogoliubov approximation, i.e., the number of particles forming the condensate must be much larger than that of forming the depleted cloud, the backreaction corrections to the standard GPE are by necessity small (for the Bogoliubov theory to apply the gas must be dilute), and consequently also the above mentioned differences from the standard linearized analogue gravity procedure (which we dub in what follows the Unruh paradigm), must be small. 
However, due to the high tunability of quantum gases, in particular with respect to the two-body coupling strength, by Feshbach resonances (cf., e.g., Refs.~\cite{Pollack,Chin}), one can potentially observe such deviations from the conventional analogue gravity paradigm in a highly controlled manner in cold quantum gas experiments.


The paper is organized as follows: In Section. \ref{sec2}, we explain the formulation of the number-conserving approach we follow throughout the paper. Section \ref{sec3} consists of the description of identifying the Lorentz invariant dynamical equation for the perturbed field. 
Section IV describes the quantum nonlinear corrections to the lowest order semiclassical background metric
and Section V elaborates on these corrections for a 
 Friedmann-Lema\^{\i}tre-Robertson-Walker metric. In Section VI
we conclude, and an Appendix contains the explicit evaluation of the drain term driving 
condensate loss for a Bose gas in a hard-walled box trap which is  
quenched from zero to finite interaction coupling. 
We finally note that we use the term {\em Unruh paradigm metric} (or simply {\em Unruh metric})  
instead of the more conventionally used ``acoustic" or ``sonic" spacetime metric, to clearly distinguish this linearized paradigm metric from the second-order metric we obtain below within our nonlinear approach to analogue gravity.

\section{number-conserving formulation of the Bose-Einstein condensed gas}\label{sec2}

\subsection{Governing field equations}
The matter wave field $\Psi$ is assumed to be governed in the standard 
$s$-wave approximation appliicable to the dilute gas 
by the following field equation ($\hbar=1$)
\begin{equation}
i\partial_t\Psi=\left(-  \frac{1}{2}\nabla^2+V+\lambda|\Psi|^2\right)\Psi,\label{fieldeq}
\end{equation} 
Here, $V$ is the external (trapping) potential, {and $\lambda>0$ is the interaction coupling from 
$s$-wave scattering between the gas particles.} 
The theory described by \eqref{fieldeq} is invariant under global U(1) transformations, which ensures the conservation of the total  particle number 
$N$ according to the Noether theorem. 
The conservation law $\partial_t\rho+\nabla \cdot {\bm J}=0$ thus holds, where 
\begin{equation}
\rho=|\Psi|^2, \qquad 
{\bm J}=  \frac{1}{2i}(\Psi^{*}\nabla\Psi-\Psi \nabla\Psi^*)
\label{totaldensitycurrent}
\end{equation}
are the system particle and current densities, respectively, and the total number of particles $N$ is given by the density integrated over all of space.

The variant of number-conserving theory we work with is based on the following 
{formal expansion of the classical field} in powers of $N^{1/2}$\footnote{Note there is a typo in 
Eq.~(2) of Ref.~\cite{baak2022number}, where $\ord(N^{-3/2})$ should be replaced by
$\mathcal{O}(N^{-1})$}  
\begin{equation}
\Psi=\phi_{\rm 0}+\chi+\phi_{2}+\mathcal{O}(N^{-1}) \label{fieldexpansion}.
\end{equation}
Here, the scaling behavior of each components taking the limit of very large $N$ 
while keeping $\lambda N=\mbox{constant}$, is given by 
\cite{schutzhold2005quantum}
\begin{equation}
	\phi_{0}=\mathcal{O}(N^{1/2}), \chi=\mathcal{O}(N^{0}),\ \mbox{and}\ \phi_{2}=\mathcal{O}(N^{-1/2}).
\end{equation}
We denote the quasiparticle (Bogoliubov) 
excitation field $\chi$ 
to keep the notation consistent with 
previous literature. 

The above formal expansion \eqref{fieldexpansion} 
relies on the assumption that the number of noncondensed particles, $\delta N$, is much less than that of those forming the condensate part, and thus posits that $N_{0} \lesssim N$, also cf.~Ref.~\cite{Lieb2002} 
for the proof that in the mathematical limit $N\rightarrow \infty$ with $\lambda N$ held fixed,  
the GPE \eqref{Eq:GrossPitaevskii} below becomes exact.
Here, $\chi$ is the fluctuation field associated with the noncondensed particles, and $\phi_{2}$ encompasses the correction of the condensate wave function due to the  backreaction of the field $\chi$ onto the condensate. 
An expansion respecting the $N$ hierarchy is necessary to properly identify the fluctuation operator and  the correction to the leading order Gross-Pitaevski\v\i~orbital, see for details below.

The basic two equations determining the dynamics of the condensate and the noncondensate part in the full number-conserving formulations are the $\ord{(N^{1/2})}$ GPE, which can be written as 
\begin{equation}
i\frac{\partial}{\partial t} \phi_{0}=\left[-  \frac{1}{2}\nabla^2+V+\lambda|\phi_{0}|^{2}
\right]
\phi_{0}.
\label{Eq:GrossPitaevskii}
\end{equation}
and the modified Bogoliubov–de Gennes (BdG) equation for the 
fluctuation field $\chi$ 
\begin{equation}
i\frac{\partial}{\partial t}\chi 
= \left(-  \frac{1}{2}\nabla^2+V+2\lambda\rho_0\right)\chi 
+\lambda\phi_{\rm 0}^2\chi^{*} 
,\label{BdG}
\end{equation}
with $\rho_0=|\phi_0|^2$.
The above equation and the corresponding Hermitian conjugate equation are the modified versions of the Bogoliubov–de Gennes equations in the non-number-conserving standard Bogoliubov approach. 
When we consider the next-order corrections to the condensate wave function due to the backreaction of the noncondensate part of the system, as was done in \cite{castin1998low, gardiner2007number}, then it can be shown that the $\mathcal{O}(N^{0})$ corrections to the leading order GPE are identically zero \cite{castin1998low}. However in the next order ${\mathcal O}(N^{-1/2})$, the standard GPE is modified due to the correction from the time-dependent backreaction of 
the noncondensate particles, as described by the fluctuation operator. 
To provide a description of the corresponding mechanism, 
we will follow \cite{schutzhold2005quantum, baak2022number}
and define a quantum-fluctuation-corrected condensate wave function 
\begin{equation}
\phi_{c} 
= \phi_{0}
+ \phi_{2}
, \label{def_phi_c}
\end{equation} 
which evolves according to the generalized GPE derived in Ref.~\cite{gardiner2007number}, 
\begin{equation}
i\partial_t\phi_{ c}=\Big(-  \frac{1}{2}\nabla^2+V+\lambda|\phi_{\rm c}|^2+2\lambda|\chi|^2\Big)\phi_{ c}+\lambda\chi^2\phi_{c}^{*}.
\label{GeneralizedGPE}
\end{equation}
This equation describes the coupled dynamics of condensed and noncondensed parts taking the dynamical corrections due to the backreaction into account. 
We note that the corrected order parameter $\phi_{ c}$ has a dominant $\ord(N^{1/2})$ term plus a subdominant correction of order $N^{-1/2}$, and for a proper formulation of backreaction, it is essential to take this correction into account. 
We pursue here the approach that backreaction is taken into account to 
this {\em minimal} order $\ord(N^{-1/2})$ in the expansion \eqref{fieldexpansion}. 
In principle, higher order corrections to the mean field $\phi_c$ in Eq.~\eqref{def_phi_c} in this formal
expansion could be considered, and thus also higher order terms would occur in the effective fields of the analogue spacetime wave equation for second-order perturbations \eqref{analoguemetric} which is  
derived below. 
We should finally also stress that we generally do not take further large $N$ approximations in the following expressions, above and beyond that of  the field expansion order taken into account for defining $\phi_c$.

To formulate the quantization of the fluctuation on top of the classical background 
constituted by $\phi_c$, we promote the fluctuation field $\chi$ and its Hermitian conjugate to be the fundamental 
quantum operators satisfying the standard bosonic commutation relations 
\begin{eqnarray} 
\left[\hat \chi ({\bm x},t),\hat \chi^\dagger({\bm x}',t)\right] = \delta({\bm x}-{\bm x}'), \nn 
\left[\hat \chi ({\bm x},t),\hat \chi ({\bm x}',t)\right] 
=\left[\hat \chi^\dagger({\bm x},t),\hat \chi^\dagger({\bm x}',t)\right] = 0. \label{bosonchi}
\end{eqnarray} 
We will then assume that the instantaneous quantum state under consideration fulfills  
\begin{equation}
\langle\hat{\chi}({\bm x},t))\rangle\coloneqq 0\,\,\,\forall\, {\bm x},t\ge 0 \label{vaccondition} 
\end{equation}
which constrains the space of states under consideration.
The initial state for the Heisenberg picture evolution is here chosen as the noninteracting $\lambda=0$ state which has zero depletion $\langle\hat\chi^\dagger\hat{\chi}\rangle =  0 \,\,(t=0)$ and $\phi_2=0$ (zero backreaction)  \cite{baak2022number}, 
and can be set up in experiment in a well controlled manner by tuning the system to the zero-crossing of a
 Fesh\-bach resonance. 
At $t=0$, we quench to a finite value of $\lambda$ and follow the subsequent evolution, maintaining the condition   \eqref{vaccondition}  
corresponding initially  to the noninteracting vacuum.

The quantization choice to regard $\hat \chi$ as the fundamental bosonic 
quantum field instead of $\hat\Psi$, does not maintain 
number conservation on the level of the full field operator of the bosonic particles: Due to $\langle \hat\Psi\rangle =\phi_c$, where however now $\phi_c$ is the backreaction-renormalized condensate mode 
in Eq.~\eqref{def_phi_c}, 
the U(1) symmetry is still broken on the quantum level, 
as in the original Bogoliubov theory \cite{Bogoliubov1947}. This choice, to promote
$\chi$ to be the ``fundamental'' quantum field,  however  
yields a set of equations amenable to a concrete solution the backreaction problem \footnote{{We note in this connection that we were unable to locate an explicit backreaction calculation following the formally rigorous and in this regard very appealing approach of  Ref.~\cite{castin1998low}, which maintains $\langle\hat \Psi\rangle=0$ 
for any fixed $N$ state.}}.   
Furthermore, and most importantly in our context, 
it has been shown in \cite{baak2022number} that with this vacuum choice for the quantized field $\hat\chi$, number conservation for condensate plus excitations is maintained, see Eqs.~\eqref{backgroundcontinuity} 
and \eqref{chicontinuity} below, and up to the order in $N^{-1/2}$  
considered.  
 
The equation governing the dynamics of the corrected mean field is then given by the expectation value of \eqref{GeneralizedGPE} 
\begin{equation}
i\partial_t\phi_{ c}=
\left[-  \frac{1}{2}\nabla^2+V+\lambda|\phi_{\rm c}|^2+2\lambda\langle\hat{\chi}^\dagger\hat{\chi}\rangle\right]\phi_{ c}+\lambda\langle\hat{\chi}^2\rangle\phi_{ c}^{*}.
\label{GeneralizedGPE2}
\end{equation}
We will not delve into the details of defining the quasiparticle operator but will rather take the (pragmatic) 
viewpoint that field quantization can be performed at least for cases where the quasiparticle vacuum is well defined, such as in the simplest, that is noninteracting case. We therefore assume that we can evaluate terms such as the anomalous density appearing in the modified GPE by solving the  $\mathcal{O}(N^{0})$ BdG equation 
as an explicit functional of the leading order GPE orbital $\phi_{ 0}$, along with the associated dependence on the spacetime coordinates $x,t$: see Appendix 
for a detailed calculation with the setup of Refs.~\cite{baak2022number,Ribeiro:2024vem}.


\subsection{Quantum backreaction}
We summarize here the formulation of the quantum backreaction problem in the 
formalism of Ref.~\cite{baak2022number}, by taking the  ${\mathcal O}(N^{-1/2})$ correction to the condensate background 
into consideration. The full system satisfies a conservation equation, 
$\partial_{t} \rho + \nabla \cdot {\bm J}=0$, where
the average density is given as  
\begin{equation}
\rho:= \langle \hat{\rho}\rangle=\langle \hat{\Psi}^{\dagger} 
\hat{\Psi} 
\rangle,
\end{equation}
and the average current is 
\begin{equation}
{\bm J}:= \langle\hat{\bm J}\rangle=  \frac{1}{2i}\left\langle\hat{\Psi}^{\dagger} 
\nabla\hat{\Psi} 
-\nabla\hat{\Psi}^{\dagger}\hat{\Psi}
\right\rangle.
\end{equation}
Utilizing the expansion of the full field operator \eqref{fieldexpansion} 
and evaluating the expressions above given the condition \eqref{vaccondition}  
 we can write the density and  current equations as
\begin{eqnarray}
\rho&=&|\phi_{c}|^{2} + \langle \hat{\chi}^{\dagger} \hat{\chi} \rangle + {\mathcal O}(N^{-1/2}), \\
{\bm J}&=&  \frac{1}{2i}\Big(\phi^{*}_{c} 
\nabla\phi_{c}-\phi_{c}\nabla\phi^{*}_{c} 
\Big) 
\nn
&+&
  \frac{1}{2i} \left\langle\hat{\chi}^{\dagger}\nabla\hat{\chi}-\nabla\hat{\chi}^{\dagger}\hat{\chi}\right\rangle + {\mathcal O}(N^{-1/2}).
\end{eqnarray}
We have ordered the various terms 
by their order in $N^{1/2}$. For a proper description of the condensate correction due to quantum fluctuations  it is useful to separate the contribution of the condensate and the noncondensate part in the full density and current as follows 
\begin{equation}
\rho_{\hat{\chi}}=\langle \hat{\chi}^{\dagger} \hat{\chi} \rangle,
\end{equation}
\vspace*{-2em}
\begin{equation}
{\bm J}_{\hat{\chi}}=  \frac{1}{2i}\left\langle\hat{\chi}^{\dagger}\nabla\hat{\chi}-\nabla\hat{\chi}^{\dagger}\hat{\chi}\right\rangle ,
\end{equation}
We then have $\rho=\rho_{c}+\rho_{\hat{\chi}}+{\mathcal O}(N^{-1/2})$ and ${\bm J}={\bm J}_{c}+{\bm J}_{\hat{\chi}}+{\mathcal O}(N^{-1/2})$, with ${\bm J}_{c}=  \frac{1}{2i}(\phi^{*}_{c} 
\nabla\phi_{c}-\phi_{c}\nabla\phi^{*}_{c} 
) $. 

Neither condensate nor depletion are separately conserved. 
In particular, we have for condensate and depletion cloud, respectively  
\cite{baak2022number,Ribeiro:2024vem} 
\begin{eqnarray}
\partial_{t}\rho_{c} + \nabla \cdot {\bm J}_{c}&=& i \lambda\Big(\phi^{2}_{c}\langle\hat{\chi}^{\dagger}\hat{\chi}^{\dagger}\rangle-\phi^{*2}_{c} \langle\hat{\chi}\hat{\chi}\rangle\Big)
\label{backgroundcontinuity}\\
\partial_{t}\rho_{\hat{\chi}} + \nabla \cdot {\bm J}_{\hat{\chi}}&=&-i \lambda\Big(\phi^{2}_{c}\langle\hat{\chi}^{\dagger}\hat{\chi}^{\dagger}\rangle-\phi^{*2}_{c} \langle\hat{\chi}\hat{\chi}\rangle\Big).\label{chicontinuity}
\end{eqnarray}
We conclude that the theory is indeed conserving up to order ${\mathcal O}(N^{0})$ for density and current. This defines the extent to which our approach is number-conserving. 
We observe that this conservation order directly corresponds to the order $\ord(N^{-1/2})$ for the 
mean-field correction $\phi_2$ in Eq.~\eqref{def_phi_c}, and is not derived by any further approximation. 
The number of particles in condensate and  in the depletion cloud of quantum excitations are not separately conserved but only their respective sums, and thus the {\em total} density and current.
The separate conservation equations for condensate and excitations acquire 
drain (source) terms which stem from backreaction.  
We furthermore emphasize that  number conservation holds here for the {\em averaged} 
quantities density and current. Fluctuations of leading relative order $\ord(N^{-1/2})$ persist, as we break number conservation on the microscopic quantum level of the elementary bosons by imposing 
$\langle\hat \Psi \rangle= \phi_c$, which corresponds to a coherent state for large $N$ with Poissonian statistics.
Note also that the prescription of quantization we employ in Eqs.~\eqref{bosonchi} implies that $\hat\Psi$ 
and $\hat \Psi^\dagger$ do {\em not} satisfy the canonical bosonic commutators. The role of the quantized bosonic fields 
is taken over by $\hat \chi,\hat\chi^\dagger$. This is akin to what happens in the original Bogoliubov approach \cite{Bogoliubov1947}.

\section{Analogue gravity in a number-conserving formalism}\label{sec3}
To understand how the propagation of fluctuations in a BEC is affected in a number-conserving description, 
we have to study how a perturbed solution of the equation (\ref{GeneralizedGPE}) behaves for sufficiently small perturbation amplitude. We first decompose via the Madelung transformation the GPE orbital as 
\begin{equation}
\phi_{ c}=\rho_{c}^{1/2}e^{i\theta_{c}},
\end{equation} 
in terms of the two real functions $\rho_{c}, \theta_{c}$. Then the modified GPE can be written in terms of the two real equations 
\begin{equation}
\partial_{t}\rho_{c}+\nabla \cdot\left(\rho_{c}\nabla\theta_{c}\right)= f_{1}(\rho_{c},\theta_{c}, {\bm x}, t),
\label{Continuity equation}
\end{equation}
\vspace*{-1em}
\begin{equation}
\begin{split}
\partial_{t}\theta_{c} +\lambda\rho_{c} 
+  \frac{1}{2}\left(\nabla\theta_{c}\right)^2 
+f_{3}(\rho_{c},\theta_{c}, {\bm x}, t) 
+V({\bm x},t)=0.
\label{Bernoulli equation}
\end{split}
\end{equation}
Here, we have defined the real functions   $f_{1}(\rho_{c},\theta_{c}, {\bm x}, t)$, and $f_{3}(\rho_{c},\theta_{c}, {\bm x}, t)$, as 
\begin{eqnarray}
\begin{split}
f_{1}(\rho_{c},\theta_{c}, {\bm x}, t) = i \lambda\Big(\phi^{2}_{c}\langle\hat{\chi}^{\dagger}\hat{\chi}^{\dagger}\rangle-\phi^{*2}_{c} \langle\hat{\chi}\hat{\chi}\rangle\Big) , \\
f_{3}(\rho_{c},\theta_{c}, {\bm x}, t) = 2\lambda\langle\hat{\chi}^{\dagger}\hat{\chi}\rangle+\frac{\lambda}{2}\Big( 
\langle\hat{\chi}\hat{\chi}\rangle e^{-2i\theta_{c}} + {\rm h.c.}
\Big).
\label{f1f2def}
\end{split}
\end{eqnarray} 
which depend on real and imaginary parts of the ``anomalous" excitation density 
$\langle\hat{\chi}\hat{\chi}\rangle$. 
We have omitted  the ``quantum pressure" term 
$-h(\rho_c) 
= -  \frac{1}{2}\rho_{c}^{-1/2}\left(\nabla^2\rho_{c}^{1/2}\right)$ on the left-hand side 
of the Bernoulli Eq.~\eqref{Bernoulli equation}; we will generally neglect density 
gradients in the background(s) 
in what follows.
Including these gradients would lead us to the concept of (nonlocal) rainbow metrics \cite{Visser:2007nx}. 
The indices 1 and 3 on $f_1$ and $f_3$ indicate their respective orders
$\ord(N^0)$ and $\ord(N^{-1})$; see also the note after Eq.~\eqref{scalings} below.

In the remainder of the paper, we will work with  \eqref{Continuity equation} and \eqref{Bernoulli equation}, which are essentially the standard fluid dynamical quasi-nonlinear equations, based on which we will explore our construction of analogue gravity in the second order nonlinear 
response regime. 
It is important to point out that the two functions 
$f_1$ and $f_3$ which contain the  deviation from the usual hydrodynamic description based on the usual 
leading order GPE are of the order $\mathcal{O}(N^{0})$ and $\mathcal{O}(N^{-1})$, respectively. 
Thus they are subdominant contributions to Eqs.~(\ref{Continuity equation} and \ref{Bernoulli equation}). 
They however encapsulate the backreaction we are interested in, and thus  
we fully keep $f_{1}, f_{3}$, assuming them to be general functions of $\rho_{c}$, and $\theta_{c}$.
Note that $f_1<0$ corresponds to the {\em loss} of particles from the condensate. 

Let us now elaborate  on two important differences from conventional analogue gravity models obtained from  the hydrodynamical equations by using the standard GPE \eqref{Eq:GrossPitaevskii}:

\textbf{(1) }If we incorporate the action of the noncondensed part in the evolution, then the condensed part of the continuity equation for  $\phi_{ c}$ is not separately conserved, and this is embedded in the drain term $f_{1}(\rho_{c},\theta_{c})$. As a result, the standard continuity equation is no longer valid, and we have to incorporate the breakdown of continuity in the formulation of the analogue gravity description based on this background. 

\textbf{(2)} Once we include the backreaction field $\phi_{2}$, then the background on which we aim to build an analogue gravity model, namely $\phi_{c}$ has two pieces, one is the usual dominant ${\mathcal O}(\sqrt{N})$ contribution $\phi_{0}$ and the other is the subdominant $\phi_{2}$, which is of order ${\mathcal O}(N^{-1/2})$. 
To encapsulate the effect of backreaction in the analogue gravity description, the perturbation expansion of the solution of the generalized GPE  around a sufficiently slowly varying background should thus also be of ${\mathcal O}(N^{-1/2})$. It is not sufficient just to consider linearized perturbations ${\mathcal O}(N^0)$ on top of a background including backreaction, as the background itself contains the field $\phi_2 = {\mathcal O}(N^{-1/2})$. 
 
In summary, we have to go beyond the standard linear fluctuation analysis to see how the backreaction of the quasiparticle excitation field changes the standard analogue gravity description \cite{datta2022analogue}. 

\subsection{Linearized fluctuations around a background}\label{linear fluctuation}
We first briefly review the basic structure of the conventional analogue gravity formulation based on linear fluctuations around a background \cite{Unruh1980, Visser:1997ux, Barcelo:2005fc}.

The fundamental two equations describing the evolution of the fluid variables $\rho_{c}$, and $\theta_{c}$, namely the (modified) continuity equation (\ref{Continuity equation}) and the Euler equation (\ref{Bernoulli equation}) are coupled first-order quasilinear partial differential equations. They have a unique solution for well-defined boundary conditions.  
The choice of ``background"  and ``perturbation" is always to some extent arbitrary.  
Conventionally one however  chooses the {\em spatiotemporally slowly} varying part of the full evolution to  constitute the background, which then also justifies our neglect of background density gradient terms 
throughout. 

We will assume a convergent power series expansion of the fluid variables ($\rho_{c}$, $\theta_{c}$) 
around the background ($\rho_{0}$, $\theta_{0}$), in the form \cite{Vandyke}
\begin{eqnarray}
\rho_{c}=\rho_0+\epsilon \delta\rho_{1}+\epsilon^{2}\delta \rho_{2}+...,
\label{expansion1}\\ 
\theta_{c}=\theta_{0}+\epsilon\delta\theta_{1}+ \epsilon^2 \delta\theta_{2}+.....
\label{expansion2}
\end{eqnarray}
Here, the scaling of the density contributions with the number of particles is as follows,
\begin{equation}
 \rho_0=\mathcal{O}(N^{1/2}), \quad \delta\rho_{1}=\mathcal{O}(N^{0}),\quad \delta\rho_2=\mathcal{O}(N^{-1/2})
 \label{scalings}
\end{equation}
and so on. For later convenience, we can thus establish a one-to-one correspondence with the subscript of each component of the density perturbation with the relevant scaling with $N^{1/2}$. For example,  
the subscript $0\rightarrow \mathcal{O}(N^{1/2})$ and similarly for other terms in descending order, i.e., an index $\alpha$ implies being of $\ord(N^{(1-\alpha)/2})$. 
Note also that $f_1=\mathcal{O}(N^{0})$ and $f_3=\mathcal{O}(N^{-1})$ defined earlier in \eqref{f1f2def} 
follow the same scaling convention, but that the phase perturbation index in \eqref{expansion2}
does not indicate a scaling with powers of $N^{1/2}$.

{We stress here for clarity that we are dealing with two expansion parameters in terms of small quantities,
where $N^{-1/2}$ is the {\em mean-field} expansion parameter and $\epsilon$ is the parameter indicating
the {\em amplitude }perturbation order of the fields.}  
 If we choose to follow the Unruh paradigm \cite{ Barcelo:2005fc, Unruh1980}, then it would  suffice to keep only ${\mathcal O}(\epsilon)$ terms, and we can study the dynamics of the perturbed phase  $\delta\theta$ around the background $\rho_0, \theta_{0}$, using that the continuity equation remains satisfied. Then it is possible to describe the dynamics of the (first-order) fluctuation field $\delta\theta_{1}$ as a massless Klein-Gordon equation in an effective curved spacetime 
\begin{equation}
\frac{1}{\sqrt{-g}}\partial_{\mu}\left(\sqrt{-g}g^{\mu\nu}\partial_{\nu}\delta\theta_{1}\right)=  0.
\end{equation}
The components of the metric $g_{\mu\nu}$ depend on the background variables $\rho_{0}, \theta_{0}$, where  the speed of sound in this background is $c^2_{s0}=\lambda\rho_{0}$. 


Importantly, this procedure implies that we are taking the background variables $\rho_0, \theta_{0}$, and the perturbed variables $\rho_{1}=\rho_0+\epsilon \delta\rho_{1}, \theta_{1}=\theta_{0}+\epsilon\delta\theta_{1}$ to be two {\em independent} 
sets of solutions of the fluid-dynamical equations, for proper boundary conditions. It is, doing so, assumed that the boundary conditions are such that expansions of the form \eqref{expansion1} and  \eqref{expansion2} are 
possible, i.e, that the corresponding series do not diverge. 
\smallskip

\subsection{Second-order fluctuations around a background}
As explained in the previous section, the orbital $\phi_{ c}$ has also a subleading contribution, and if we want to see how the effect of such a subleading contribution changes the conventional Unruh paradigm, then it is essential to consider the series expansion (\ref{expansion1}) at least up to $\delta\rho_2$. However, it is also important to understand how the terms $\mathcal{O}(N^{0})$, i.e. $\delta\rho_{1}$, change the previous background $\rho_{0}, \theta_{0}$, and if it is at all possible to provide an analogue gravity description with respect to the new background \cite{datta2022analogue}. 

Let us first assume perturbation amplitude expansions as in Eqs.~\eqref{expansion1} and \eqref{expansion2}) are possible, up to the indicated quadratic order. Then from the modified continuity equation (\ref{Continuity equation}), and the Euler equation \ref{Bernoulli equation}, we obtain the following set of equations:
\begin{widetext}
\begin{eqnarray}
\partial_{t}(\rho_{0}+\epsilon\delta\rho_{1})+\nabla\cdot\Big((\rho_{0}+\epsilon\delta\rho_{1})(\nabla\theta_{0}+\epsilon\nabla\delta\theta_{1})\Big)=f_{1}(\rho_{0}, \theta_{0})
+ \epsilon \dot f_1 
\delta\rho_{1}+\epsilon
f_1^\prime
\delta\theta_{1} 
+\epsilon^2
\dot f_1^{\prime}
\delta\rho_{1}\delta\theta_{1},
\label{background1}\\ 
	\partial_{t}(\theta_{0}+\epsilon\delta\theta_{1})
	+\lambda(\rho_{0}+\epsilon\delta\rho_{1})+  \frac{1}{2}(\nabla(\theta_{0}+\epsilon\delta\theta_{1}))^2+V 
	+f_{3}(\rho_{0}, \theta_{0})+\epsilon \dot f_3
	\delta\rho_{1}+\epsilon
	f_3^\prime 
	\delta\theta_{1} 
	+\epsilon^2 \dot f^\prime _3 
	\delta\rho_{1}\delta\theta_{1}=0.
	\label{background2}
	\end{eqnarray}
\end{widetext} 
where we introduced the abbreviating notation that 
partial derivatives with respect to $\rho_{c}$ and $\theta_{c}$, taken at 
$\rho_c=\rho_0$ and $\theta_c= \theta_0$, are represented as an overdot  for $\rho_{c}$ derivatives 
and as a prime for $\theta_{c}$ derivatives. Note in this regard that an overdot thus implies a scaling factor $1/N$ to leading order. 
As we will argue later in detail, the two above equations 
define the new background  on which we will consider the propagation of nonlinear fluctuations. Then the 
new equations governing the perturbations are 
\begin{widetext}
	\begin{eqnarray}
	\epsilon^2\partial_{t}\delta\rho_2+\nabla\cdot\Big(\epsilon^2\delta\rho
	_2(\nabla\theta_{0}+\epsilon\nabla\delta\theta_{1})+\epsilon^2(\rho_{0}+\epsilon\delta\rho_{1})\nabla\delta\theta_2\Big)=\epsilon^2 \ddot f_1 
	\delta\rho_2+\epsilon^2 f_1^{\prime\prime} 
	\delta\theta_2,
	\label{perturbation1}\\
\epsilon^2\partial_{t}\delta\theta_2+\epsilon^2\lambda\delta\rho_2+\epsilon^2 \nabla(\theta_{0}+\epsilon\delta\theta_{1})\cdot\nabla\delta\theta_2+\epsilon^2
\ddot f_3 
\delta\rho_2+\epsilon^2
f_3^{\prime\prime}
\delta\theta_2=0.
\label{perturbation2}
\end{eqnarray}
\end{widetext}
First of all, here we have, in accordance with the subleading contribution to the 
background including backreaction, used an {\em inhomogeneous separation} of effective contributions, 
controlled by the separation parameter $\epsilon$ ({\em inhomogeneity} in this parameter), to transform the initial fluid-dynamical equations for the fields into a set of four equations for ``background," Eqs.~\eqref{background1} and  \eqref{background2}
and ``perturbation," \eqref{perturbation1}  and  \eqref{perturbation2}, cf.~ the corresponding procedures in Ref.~\cite{Nayfeh2008perturbation}. As we will see, this is mandatory to obtain an analogue gravity  description of these nonlinear perturbation equations.  

Let us consider the following combinations of variables, 
\begin{eqnarray}
\rho_{b}=\rho_{0}+\epsilon\delta\rho_{1},\quad
\theta_{b}=\theta_{0}+\epsilon\delta\theta_{1},\label{newbackground}\\
{\rho_c} = \rho_{0}+\epsilon\delta\rho_{1}+\epsilon^2\delta\rho_2, \quad
{\theta_c}= \theta_{0}+\epsilon\delta\theta_{1}+\epsilon^2\delta\theta_2 ,
\end{eqnarray} 
where we truncated in the second line the series in Eqs.~\eqref{expansion1} and \eqref{expansion2} after 
second order in the fluid variables $\delta\rho_2$ and $\delta\theta_2$. 
We assume that these combinations 
satisfy the dynamical fluid Eqs.~\eqref{Continuity equation} and \eqref{Bernoulli equation}) for two  
different boundary conditions, such that the expansions above are convergent, which in turn requires 
$\epsilon^2\delta\rho_2 \ll \epsilon\delta\rho_1\ll\rho_0$. This also implies that we are not separately demanding $\rho_{0}, \theta_{0}$ to follow any dynamical equations, at least formally. Then the condition on $\rho_{0}, \theta_{0}$ is simply that they are two sufficiently slowly varying ``background" functions, which are not necessarily solutions to the fluid dynamical equations, but can be chosen, for example,
 according to the ease of interpretation of experiments. Namely, such that the first and second order perturbative corrections on top of the background satisfy the fluid-dynamical equations.  
 
We can consider of course also fluctuations higher than of quadratic order and, in a similar {fashion}, 
demand that we can suitably define a background for the next-order density and phase fluctuations. 
In a minimally self-consistent analogue gravity model, as a consequence of the backreaction due to enforcing number conservation, the perturbed fluid variables have however to be {\em at least} of 
quadratic order, i.e., should scale here as $\mathcal{O}(N^{-1/2})$, to be properly treated as a perturbation.  

Taking into account the variables ($\rho_{b}, \theta_{b}$) of Eq.~\eqref{newbackground}  
defining the new background, 
we can substitute $\delta\rho_2$ from (\ref{perturbation2}) into (\ref{perturbation1}) to obtain 
\begin{widetext}
	\begin{multline}
	\label{thetadynamics}
	\partial_{t}\Big(\frac{1}{\tilde{\lambda}}\Big[\partial_{t}\delta\theta_{2}+{\bm v}_b\cdot\nabla\delta\theta_{2}+f^{\prime\prime}_{3}\delta\theta_{2}\Big]\Big)
	+ 
	\nabla \cdot\Bigg(\frac{1}{\tilde{\lambda}}{\bm v}_b\Big[\partial_{t}\delta\theta_{2}+{\bm v}_b\cdot\nabla\delta\theta_{2}+f^{\prime\prime}_{3}\delta\theta_{2}\Big]-\rho_{b}\nabla\delta\theta_{2}\Bigg)  \\
	-\frac{\ddot{f}_{1}}{\tilde{\lambda}}\Big(\partial_{t}\delta\theta_{2}+{\bm v}_b\cdot\nabla\delta\theta_{2}+f^{\prime\prime}_{3}\delta\theta_{2}\Big)+f^{\prime\prime}_{1}\delta\theta_{2}=0.
	\end{multline}
\end{widetext}
We have defined the background velocity 
\begin{equation} 
{\bm v}_{b}=\nabla(\theta_{0}+\epsilon \delta\theta_{1}), 
\end{equation} 
and an effective coupling constant 
 \begin{equation}
\label{tildeU}
\tilde{\lambda}=\lambda 
+\ddot f_3.
\end{equation}

\subsection{Lorentz-invariant wave equation for second-order perturbations} 
As the continuity equations are not satisfied separately, e.g., the condensate loses particles, there are extra terms present  that indicate that the dynamics of the perturbation field $\delta\theta_{2}$ can not be written simply as a Lorentz-invariant curved spacetime equation. 

However, it is possible to obtain a Lorentz invariant wave equation for the dynamics of the perturbation field.
This comes at the price of introducing additional fields, though, as we will now discuss in detail. 
We obtain from Eq.~\eqref{thetadynamics} (without any further approximation) the following equation of motion
\begin{multline}
\frac{1}{\sqrt{-\mbox{\sf g}}}\mathcal{D}_{\mu}\left(\sqrt{-\mbox{\sf g}}\mbox{\sf g}^{\mu\nu}\mathcal{D}_{\nu}\delta\theta_{2}\right)
\\
-\mbox{\sf g}^{\mu\nu}\mathcal{A}_{\mu}\mathcal{D}_{\nu}\delta\theta_{2} -\mathcal{B}^{\mu}\mathcal{D}_{\mu}\delta\theta_{2}+ 
\mathcal{M}^2\delta\theta_{2}=  0.
\label{analoguemetric}
\end{multline}
Here, we have defined the modified derivative operator as 
\begin{equation}
\mathcal{D}_{\mu} =\partial_{\mu}+\mathcal{A}_{\mu}. \label{defA}
\end{equation}
{Note that $A_\mu$ does not assume the role of a gauge field here and is defined such that it is real; 
see 
Eq.~\eqref{vectorA} below.} 

We can identify the analogue metric components 
as follows 
\begin{eqnarray}
\mbox{\sf g}_{tt}&=& -\sqrt{\frac{\rho_{b}}{\tilde{\lambda}}}\left(\tilde{\lambda}\rho_{b}-{\bm v}_b^{2}\right),
\label{metrictt}\\ 
\mbox{\sf g}_{ti}&=& -\sqrt{\frac{\rho_{b}}{\tilde{\lambda}}}{\bm v}_{bi},\qquad 
 \mbox{\sf g}_{ij}= \sqrt{\frac{\rho_{b}}{\tilde{\lambda}}}\delta_{ij},
\label{metricti}
\end{eqnarray}
where the metric determinant is given by 
\begin{equation}
\sqrt{-\mbox{\sf g}}= \sqrt{\frac{\rho_{b}^{3}}{\tilde{\lambda}}}. 
\end{equation} 
Here, 
the index $i$ represents the spatial coordinates. 

We have defined in Eqs.~\eqref{analoguemetric} and \eqref{defA} the components of a 
vector field 
\begin{equation}
\mathcal{A}^{\mu}=\Big(\frac{f^{\prime\prime}_{3}}{\sqrt{\mbox{$-$\sf g}}\tilde{\lambda}}, \frac{f^{\prime\prime}_{3}{\bm v}_b}{\sqrt{\mbox{$-$\sf g}}\tilde{\lambda}}\Big),
\label{vectorA}
\end{equation}
and the components of a second vector field $\mathcal{B}^{\mu}$ as
\begin{equation}
\mathcal{B}^{\mu}=\Big(\frac{\ddot{f}_{1}}{\sqrt{\mbox{$-$\sf g}}\tilde{\lambda}}, \frac{\ddot{f}_{1}{\bm v}_b}{\sqrt{\mbox{$-$\sf g}}\tilde{\lambda}} \Big).
\label{vectorB}
\end{equation}
Finally,  the effective mass of the scalar field is identified as follows 
\begin{equation}
\mathcal{M}^2=\frac{1}{\sqrt{\mbox{$-$\sf g}}}\Big[-\frac{2\ddot{f}_{1}f^{\prime\prime}_{3}}{\tilde{\lambda}}+f^{\prime\prime}_{1}\Big].
\label{mass}
\end{equation}
Note that spacetime indices are raised and lowered throughout with the full metric 
$\mbox{\sf g}_{\mu\nu}$, not with the Unruh metric $g_{\mu\nu}$. 

As can be readily concluded from the above Eqs.~(\ref{vectorA}, \eqref{vectorB}, and \eqref{mass}), the corresponding fields are identically zero when the two functions $f_{1}, f_{3}$ vanish, producing the result of the Unruh paradigm in terms of an effective metric only then also to second order in perturbations.  

In summary, the dynamics of the nonlinear phase fluctuations 
can be written, to second order in the perturbations, as a generalized massive Klein-Gordon equation in a curved spacetime, coupled with two emergent vector fields, all with respect to the background 
$\rho_{b}, {\bm v}_b$.

The contributions of the final two (vector field)
terms in the dynamics of the scalar field are subdominant (are of lower order in $N^{-1/2}$) 
when compared to the first two terms. The second-order emergent metric $\mbox{\sf g}_{\mu\nu}$ 
includes a term $h_{\mu\nu}$ which signifies the deviation from the Unruh paradigm metric $g_{\mu\nu}$. 
The $N$-scalings of the various terms are, to leading order, as follows  
\begin{equation}
\begin{split}
\mbox{\sf g}_{\mu\nu}=g_{\mu\nu} [\mathcal{O}(N)]+ h_{\mu\nu}  [\mathcal{O}(N^0)],\\
A_{\mu}=[\mathcal{O}(N^{-1})],\\
B_{\mu}=[\mathcal{O}(N^{-2})],\\
\mathcal{M}^2=[\mathcal{O}(N^{-1})], 
\label{scaling}
\end{split}
\end{equation} 
where $g_{\mu\nu} [\mathcal{O}(N)]$ is the conventional Unruh paradigm analogue metric.
Hence to leading order in the large $N$ expansion, corrections to the Unruh metric dominate
the quantum nonlinear effects in the Lorentz-invariant wave equation \eqref{analoguemetric}. 

In our context of the BEC-based analogue gravity model incorporating number conservation, the above corrections to the Unruh paradigm can be readily understood by considering a system where the interaction is turned on suddenly in an initially noninteracting gas \cite{baak2022number}, also see the Appendix. Then, at sufficiently early time, when the number of depleted particles is very small, the correction due to backreaction is also negligible, and the Unruh paradigm metric is an excellent description. However, as the system evolves, the number of depleted particles is no longer negligible, and 
the corrected metric will afford a better description. 

The equation \eqref{analoguemetric} is \textit{not} invariant under time reversal $t \rightarrow -t$. The presence of a first-order derivative in time breaks the time reversal symmetry, which in turn can be traced back to the existence of particle exchange between the condensed and the noncondensed parts of the system, 
which is fully respected in our construction of a number-conserving theory. The manifestation of the breaking
of time reversal is a complex quasiparticle frequency, cf.~Eq.~\eqref{dispersion} below.
\vspace*{1em}

\subsection{{Comparison to other approaches}}
At this point, it is appropriate to draw a comparison between the formulation we have adopted here and 
in particular the one proposed  in  the classical analysis of \cite{datta2022analogue}, where the propagation of a general nonlinear wave in a fluid medium described by the coupled continuity and Euler equations was considered from the viewpoint of analogue gravity. 

It was shown in  \cite{datta2022analogue} that 
an effective metric can be derived also when the intrinsic nonlinearity of a (classical perfect) fluid is taken into account. This description however breaks down after a certain time scale due to the inherent nonlinear nature of the fluid motion {and the consequent creation of shocks.} It is nevertheless possible to obtain a spacetime metric description provided we take the backreaction of the nonlinear motion into account 
and consequently modify the background itself. In the present paper, we have taken a weaker restriction on the nature of nonlinearity in  the fluid under consideration. To be more specific, distinct from \cite{datta2022analogue}, we have assumed that the perturbation expansion in \eqref{expansion1} is convergent, i.e., the successive terms in this expansion are gradually of sufficiently 
 lower order than the previous one, an assumption that was not a priori taken in \cite{datta2022analogue}. 
 Also our analysis  will break down after a certain timescale at which the fully nonperturbative nature of the wave propagation can no longer be neglected. The exact relation between this timescale, and the failure of the Bogoliubov approximation, i.e., when the number of condensate particles becomes of order the number of  noncondensed particles, should be an interesting topic to pursue in the future.

\begin{widetext}
\begin{center}
\begin{table}[t]
\begin{tabular}{|c|c|c|c|}
\hline
\textbf{Approach} & \textbf{Background considered}& \textbf{Number-conserving} & \textbf{References} \\\hline
  Leading-order GPE~\eqref{Eq:GrossPitaevskii}
&~~ Linearized on top of $(\rho_{0}, {\bm v}_{ 0}) $& No  & \cite{Barcelo:2000tg,Barcelo:2003wu,Fedichev:2003bv}
\\\hline
Number-conserving ansatz \eqref{fieldexpansion} & Full fluid density and velocity\footnote{In Refs.~\cite{schutzhold2005quantum},and \cite{baak2022number} the importance of full number conservation was pointed out. However, no consequences for the analogue metric\\ \noindent to any given order in the large $N$ expansion  
were considered.} 
& Yes & \cite{schutzhold2005quantum} \\\hline  
Nonlinear waves in classical fluid & Nonperturbative fluctuation (redefined background)\footnote{The formulation in \cite{datta2022analogue} is for a generic barotropic, non-viscous, and irrotational fluid, with a BEC as a particular case. However, the redefined background in \cite{datta2022analogue} is not the 
$(\rho_{b}, {\bm v}_b)$ employed here. } & N/A & \cite{datta2022analogue}
\\\hline
Corrected GPE \eqref{GeneralizedGPE2}& Second-order on background $(\rho_{b}, {\bm v}_b)$ & Yes & 
{\cite{baak2022number}}, This work
\\\hline
\end{tabular}
\caption{A table comparing formulations used to construct analogue gravity models for quantum and classical 
fluids.}  
\label{backgroundtable}
\end{table}
\end{center} 
\end{widetext}

 In the context of the present work, as there is a well defined separation in powers of $N^{-1/2}$, it is
 essential to explore in detail how the dynamics of nonlinear perturbations changes due to backreaction. In fact, as the backreaction of the noncondensed part is $\mathcal{O}(N^{-1/2})$, a fully nonlinear analysis as in \cite{datta2022analogue} might not be sensitive (in a controlled manner) to this correction.

 We have summarized the features of various approaches to analogue gravity {in 
quantum and classical fluids} which have been taken since its inauguration, in table \ref{backgroundtable}, together with a few selected references. The combination of generality and degree of control possible within our {quantum many-body approach} (which can be extended to higher orders of powers of $N^{-1/2}$) becomes apparent. {Most importantly, we provide 
the first spacetime metric description of the propagation of second-order perturbations in a quantum fluid, 
within a number-conserving formulation of backreaction.} 

\subsection{Dispersion relation}
To understand the nature of the dispersion relation, let us consider a plane 
	wave solution of \eqref{analoguemetric} propagating in $x$ direction,  
\begin{equation}
\delta\theta_{2}\propto \exp[ i(kx-\omega t)],
\end{equation}  
where 
the frequency $\omega$ can in general be complex. Substituting this ansatz back into Eq.~\eqref{thetadynamics}, and assuming that spatiotemporal variations of the background are negligible, we have 
\begin{multline}
\omega=v_{bx}k+ \frac{1}{2}\left[ i(\ddot{f}_{1}-f^{\prime\prime}_{3}) \right.\\
\left.\pm \sqrt{4(\tilde{\lambda}f^{\prime\prime}_{1}-\ddot{f}_{1}f^{\prime\prime}_{3}+c^{2}_{s,b}k^2)-(\ddot{f}_{1}-f^{\prime\prime}_{3})^2}\right].
\label{dispersion} 
\end{multline}
Here, we have defined the effective sound velocity of the background ($\rho_{b}$, $\theta_{b}$) to be
$c_{s,b}^2=\tilde{\lambda}\rho_{b}$. As expected, when $f_{1}=f_{3}=0$, this dispersion relation reduces to the standard Bogoliubov dispersion. 
The appearance of 
a generally complex frequency is a direct manifestation of backreaction, 
causing the condensate particle number to be not separately conserved. The exact rate at which this occurs, and the corresponding correction on the leading-order Bogoliubov dispersion depends on the relevant field configurations leading to the solutions of the BdG equation \eqref{BdG}. 

In the Appendix, we discuss a specific time dependent configuration for a BEC contained in a box potential, where the quantization of the depletion field $\hat{\chi}$ 
can be performed analytically \cite{baak2022number}. Consequently, it is possible to calculate the anomalous correlators appearing in the definitions of $f_{1}$ and $f_{3}$.  As we explain in the Appendix 
in detail, it can be concluded that for early times, after switching on interactions from an initially noninteracting state, the particle loss from the condensate, quantified by the function $f_{1}$, is quadratic in time, 
and this  loss is in such a setup generic in the sense that it is independent of the system size, at least for sufficiently early times. As a result, the leading order correction to the Unruh paradigm metric is also generic and governed by $h_{\mu\nu}$ in \eqref{scaling}, until the Bogoliubov approximation itself breaks down.

\section{Quantum nonlinear spacetime metric}\label{sec4}
Even though the form of the analogue  metric that couples to the second-order fluctuation field, $\mbox{\sf g}_{\mu\nu}$ is formally identical to that of a metric corresponding to the linear regime, the exact functional dependence on the underlying fluid variables is different, due to the fact that now the definition of background has changed. In this section we will explicitly calculate the components of the metric and will see how they differ from the standard Unruh paradigm. In the remainder of the paper we will denote the emergent nonlinear background as \textit{\textbf{background-II}} (variables $\rho_{b}$, $\theta_{b}$), while the 
background related to the strict linearization leading to the Unruh paradigm will be denoted without italics and boldface as background-I (variables $\rho_{0}$, $\theta_{0}$). 
For both backgrounds we demand, as usual, that only perturbations which vary much faster than the 
background are considered. In particular, the relevant length scales of  \textit{\textbf{background-II}} must always be much less than those of background-I for them to be meaningfully separable. 

\subsection{Metric coupled to second-order fluctuations}
Expanding the new metric components given by Eqs.~\eqref{metrictt} 
and \eqref{metricti}, and using the fluid variables $\rho_{b}, \theta_{b}$ to define the new \textit{\textbf{background-II}} we obtain 
\begin{multline}
\mbox{\sf g}_{tt}=g_{tt}-\sqrt{\frac{\rho_{0}}{\lambda}}\Big[\epsilon \lambda\delta\rho_{1}-2\epsilon {\bm v}_0 \cdot {\bm v}_1-\epsilon^2 {\bm v_1}^2 \\
+\epsilon\frac{\delta\rho_{1}}{2\rho_{0}}\big(c_{s0}^2-{\bm v}^2_{0}+\epsilon \lambda\delta\rho_{1}-2\epsilon {\bm v}_0\cdot {\bm v}_1\big)-\epsilon^2\frac{\delta\rho^2_{1}}{8\rho^2_{0}}\big(c_{s0}^2-{\bm v}^2_{0}\big)\Big], \label{gtt}
\end{multline}
\vspace*{-1.5em}
\begin{equation}
\mbox{\sf g}_{ti}=g_{ti}-\sqrt{\frac{\rho_{0}}{\lambda}}\Big[\epsilon   {\bm v}_{1i}+\epsilon\frac{\delta\rho_{1}}{2\rho_{0}}\big({\bm v}_0+\epsilon {\bm v}_{1}\big)_{i}-\epsilon^2\frac{\delta\rho^2_{1}}{8\rho^2_{0}}{\bm v}_{0i}\Big].
\label{gti}
\end{equation}
\vspace*{-1.5em}
\begin{equation}
\mbox{\sf g}_{ij}=g_{ij}+\sqrt{\frac{\rho_{0}}{\lambda}}\Big[\epsilon\frac{\delta\rho_{1}}{2\rho_{0}}-\epsilon^2 \frac{\delta\rho_{1}^2}{8\rho_{0}^2}\Big]\delta_{ij}.
\label{gij}
\end{equation}
Here, we have defined the first-order correction to the velocity field as ${\bm v}_1=\nabla\delta\theta_{1}$, and retained terms up to $\mathcal{O}(\epsilon^2)$ in the first-order perturbed fields $\delta\rho_{1}$ and $\delta\theta_{1}$, which represents the working order for the new \textit{\textbf{background-II}}. 

The standard linearized Unruh paradigm metric has components  
\begin{multline}
g_{tt}= -\sqrt{\frac{\rho_{0}}{\lambda}}\left(c_{s0}^2-{\bm v}_0^{2}\right), \qquad 
g_{ti}= -\sqrt{\frac{\rho_{0}}{\lambda}}{\bm v}_{0i},  \\
g_{ij}= \sqrt{\frac{\rho_{0}}{\lambda}}\delta_{ij}.
\end{multline}
The fact that the leading-order contribution to the effective metric that describes the dynamics of the second-order fluctuation field $\delta\theta_{2}$ is given by the Unruh metric components related to 
background-I indicates that the spacetime metrics corresponding to the two backgrounds are related by 
\begin{equation}
\mbox{\sf g}_{\mu\nu}=g_{\mu\nu}+h_{\mu\nu},
\end{equation}  
where $h_{\mu\nu}$ can be thought of as {\em emergent}, as it is  
due to the inherent nonlinearity of \textit{\textbf{background-II}}.
From Eq.~\eqref{scaling}, we conclude that the leading-order corrections come from the nonlinear contribution to the modified metric ($h_{\mu\nu}$) with respect to the \textit{\textbf{background-II}}. In fact, 
as might be expected on general physical grounds, this correction is of the same order as the density of the depletion cloud $\rho_{\hat{\chi}}=[\mathcal{O}(N^0)]$. 

\subsection{Dynamics of the background field generating curved spacetime}
To understand how changing the form of background affects the analogue gravity description in our formalism, we start with a simple example. Let us consider a uniform and static fluid medium in one spatial dimension, described by the variables $\rho_{0}, \theta_{0}$. Then the  metric for {\em linear} sound propagation is Minkowski, which corresponds to background-I. Now we consider the propagation of a nonlinear perturbation; then, as we described in the above, it is possible to obtain a Lorentz-invariant description only when we couple the second-order field $\delta\theta_{2}$ with the \textit{\textbf{background-II}}, which now however does not represent a simple Minkowski spacetime. 
The components of the tensor $h_{\mu\nu}$ in this case can be obtained by setting ${\bm v}_0=0$ and $\rho_{0}=$\,constant in Eqs.~\eqref{gtt}, \eqref{gij}, and \eqref{gti}. It is also important to note that we have imposed an extra condition on the fluid variables, namely that the initial configuration $\rho_{0}, \theta_{0}$ satisfies the fluid-dynamical equations separately, which however is not necessary to obtain the metric for second-order fluctuation fields. This relates to employing distinct boundary and initial conditions, which are imposed on the quasi-linear fluid dynamical equations we are dealing with, and again emphasizes how the distinction between the background and the fluctuation fields is to some extent arbitrary and can thus also change an analogue gravity description significantly.

The dynamical equation satisfied by this scalar field generating curved spacetime can be obtained from the two equations that define \textit{\textbf{background-II}}. Substituting $\delta\rho_{1}$ from Eq.~\eqref{background2} into Eq.~\eqref{background1}, we obtain the nonlinear equation of motion for the field $\delta\theta_{1}$ as, cf.~Ref.~\cite{datta2022analogue}, 
\begin{widetext}
\begin{equation}
\begin{split}
\partial_t\Big[-\frac{1}{\bar{\lambda}}\Big(\partial_t\delta\theta_{1}+{\bm v}_0\cdot\nabla\delta\theta_{1}+\frac{\epsilon}{2}(\nabla\delta\theta_{1})^2+
f_3^\prime 
\delta\theta_{1}\Big)\Big]+
\nabla\cdot\Big[\rho_{0}\nabla\delta\theta_{1}
-\frac{ {\bm v}_0}{\bar{\lambda}}\Big(\partial_t\delta\theta_{1}
+{\bm v}_0\cdot\nabla\delta\theta_{1}
+\frac{\epsilon}{2}(\nabla\delta\theta_{1})^2
+f_3^\prime 
\delta\theta_{1}\Big)\\
-\frac{\epsilon}{\bar{\lambda}}(\nabla\delta\theta_{1})\Big(\partial_t\delta\theta_{1}+{\bm v}_0\cdot\nabla\delta\theta_{1}+\frac{\epsilon}{2}(\nabla\delta\theta_{1})^2
+f_3^\prime 
\delta\theta_{1}\Big)\Big]\\
+\frac{1}{\bar{\lambda}} \dot f_1
\Big(\partial_t\delta\theta_{1}+{\bm v}_0\cdot\nabla\delta\theta_{1}+\frac{\epsilon}{2}(\nabla\delta\theta_{1})^2
+f_3^\prime 
\delta\theta_{1}\Big)
- f_1^\prime 
\delta\theta_{1} 
-\frac{\epsilon}{\bar{\lambda}} \dot f^\prime_1
\Big(\partial_t\delta\theta_{1}+{\bm v}_0\cdot\nabla\delta\theta_{1}+\frac{\epsilon}{2}(\nabla\delta\theta_{1})^2
+f_3^\prime 
\delta\theta_{1}\Big)\delta\theta_{1}=0.
\label{gravitydynamics}
\end{split}
\vspace{-6em}
\end{equation}
\end{widetext}
We have defined  another modified coupling as follows 
\begin{equation} 
\bar{\lambda}=\lambda 
+\dot f_3
\end{equation}
{neglecting a sub-leading correction of $\mathcal{O}(\epsilon^2)$}. 
 This should be compared with its 
 counterpart  $\tilde\lambda$  in Eq.~\eqref{tildeU}.  

The above highly nonlinear equation \eqref{gravitydynamics} describes the dynamics of 
the spacetime \textit{\textbf{background-II}} due to the intrinsically nonlinear nature of quantum fluctuations. 
We apply it in the next section to a cosmological scenario. 

\section{Quantum nonlinear corrections to a cosmological background}

\subsection{Friedmann-Lema\^{i}tre-Robertson-Walker background}

We will now discuss in more detail how the quantum nonlinear effects coming from the backreaction of the quantized field $\hat{\chi}$ change the leading-order Unruh paradigm metric, which governs 
perturbations to linear order.
To exemplify our arguments in the preceding sections we will consider a solution of the GPE for $\phi_{ 0}$, such that the leading-order $\mathcal{O}(N)$ density $\rho_{0}$ is constant and the condensate is at rest. 
This produces a simple analogue of the Friedmann-Lema\^{i}tre-Robertson-Walker (FLRW) metric of cosmology when $c_{s0}=c_{s0}(t)$, cf. Ref.~\cite{Jain2007}, where the time dependence of the sound speed is implemented by using a Feshbach resonance sweep \cite{Chin} such that $\lambda=\lambda(t)$, 
with the constant density enforced by a hard-walled trap of infinite height.

Then the Unruh paradigm metric, as seen by the first order fluctuation field $\delta\rho_{1}, \delta\theta_{1}$, is given by 
\begin{equation}
g^{\rm FLRW}_{\mu\nu}=
\sqrt{\frac{\rho_0}{\lambda(t)}}
\begin{bmatrix}
-{c_{s0}^2(t)}& 0 & 0 & 0
\\
0 & 1 & 0 & 0
\\0 & 0 &  1 & 0\\
0 & 0 & 0 & 1
\end{bmatrix},
\label{linearflrw}
\end{equation}
which is spacetime-conformally flat, according to the usual 
assumption that the (analogue) universe is isotropic and homogeneous at the cosmological scale. 
The purely spatial part of the metric is by itself also flat because the ultracold quantum gas is probed in a nonrelativistic laboratory, i.e. in Euclidean space. 
In this model, a decreasing $\lambda(t)$ will produce a spacetime corresponding to an expanding universe, and vice versa. The metric of \eqref{linearflrw} refers to the background-I, which consists of fluid variables $\rho_{0}, \theta_{0}$, that is, solutions of the leading-order GPE \eqref{Eq:GrossPitaevskii}.

However, as we explained in sections \ref{sec3} and \ref{sec4}, the spacetime metric as seen by 
the second-order fluctuation field $\delta\theta_{2}$ is not the simple FLRW metric of \eqref{linearflrw}, as the field $\delta\theta_{2}$ does not couple to the background-I, rather it couples to the redefined \textit{\textbf{background-II}} defined in terms of the fluid variables $\rho_{b}$, $\theta_{b}$. As a result, the leading-order FLRW type Unruh metric will receive, in that perturbative order, subleading corrections of $\mathcal{O}(N^0)$, the general form of which is quoted in Eqs.~\eqref{gtt}, \eqref{gti}, and \eqref{gij}, and the dynamics of which is governed by \eqref{gravitydynamics}.

In the next subsection, we study the form of the quantum fluctuation renormalized 
metric $\mbox{\sf g}_{\mu\nu}$ 
when the first-order fluctuation field  $\delta\theta_{1}$ perceives 
a Minkowski or FLRW Unruh paradigm metric $g_{\mu\nu}$. 

\subsection{Renormalized FLRW 
metric}

As stated, the background-I ($\rho_{0}$, $\theta_{0}$), corresponding to the first order fluctuation field ($\delta$$\theta_{1}$) is a FLRW metric if the sound speed in BEC is time dependent. On the other hand, the metric as seen by the second-order fluctuation field $\delta\theta_{2}$, which only couples with the \textit{\textbf{background-II}}, a solution of the corrected GP equation \eqref{GeneralizedGPE2}, is not the simple FLRW metric of the form \eqref{linearflrw}. 
The latter changes due to backreaction, from Eqs.~\eqref{gtt}--\eqref{gij}, as follows    \pagebreak 
\begin{multline}
\mbox{\sf g}_{tt}=g^{\rm FLRW}_{tt}-\sqrt{\frac{\rho_{0}}{\lambda(t)}}\Big[\Big(\epsilon \lambda(t)\delta\rho_{1}-\epsilon^2 {\bm v}^{2}_{1}\Big)\\
+\epsilon\frac{\delta\rho_{1}}{2\rho_{0}}\Big(\lambda(t)\rho_{0}
+\epsilon \lambda(t)\delta\rho_{1}\Big)-\epsilon^2\lambda(t)\rho_{0}\frac{\delta\rho^2_{1}}{8\rho^2_{0}}\Big]\\
=-\sqrt{\frac{\rho_{0}}{\lambda(t)}}\Big[\lambda(t)\rho_0+\frac{3}{2}\epsilon \lambda(t)\delta\rho_{1}+\frac{3}{8}\epsilon^2\lambda(t)\frac{\delta\rho^2_{1}}{8\rho_{0}}\Big],
\label{gttflrw}
\end{multline}
\vspace*{-2em}
\begin{eqnarray}
\mbox{\sf g}_{ti}&=&g^{\rm FLRW}_{ti}-\sqrt{\frac{\rho_{0}}{\lambda(t)}}\Big[\epsilon   {\bm v}_{1i}+\epsilon^2\frac{\delta\rho_{1}}{2\rho_{0}} {\bm v}_{1i}\Big],
\label{gtiflrw}\\
\mbox{\sf g}_{ij}&=&g^{\rm FLRW}_{ij}+\sqrt{\frac{\rho_{0}}{\lambda(t)}}\Big[\epsilon\frac{\delta\rho_{1}}{2\rho_{0}}-\epsilon^2 \frac{\delta\rho_{1}^2}{8\rho_{0}^2}\Big]\delta_{ij}\nn
&=&\sqrt{\frac{\rho_{0}}{\lambda(t)}}\Big[1+\epsilon\frac{\delta\rho_{1}}{2\rho_{0}}-\epsilon^2 \frac{\delta\rho_{1}^2}{8\rho_{0}^2}\Big]\delta_{ij}.
\label{gijflrw}
\end{eqnarray}

Here the boundary conditions are chosen such that the background-I ($\rho_{0},{\bm v}_0$) is a solution of the Euler and continuity equations for these variables. 
The dynamics of the background field $\delta\theta_{1}$ is governed by the nonlinear equation \eqref{gravitydynamics}, which here takes the somewhat simplified but still rather involved form
\vspace*{2em}
\begin{widetext}
	\begin{equation}
	\begin{split}
	\partial_t\Big[-\frac{1}{\bar{\lambda} (t)}\big(\partial_t\delta\theta_{1}+\frac{\epsilon}{2}(\nabla\delta\theta_{1})^2
	+f_3^\prime 
	\delta\theta_{1}\big)\Big]+
	\nabla\cdot\Big[\rho_{0}\nabla\delta\theta_{1}
-\frac{\epsilon}{\bar{\lambda}(t)}(\nabla\delta\theta_{1})\Big(\partial_t\delta\theta_{1}+\frac{\epsilon}{2}(\nabla\delta\theta_{1})^2
+f_3^\prime
\delta\theta_{1}\Big)\Big]\\
	+\frac{1}{\bar{\lambda}(t)} \dot f_1 
	\Big(\partial_t\delta\theta_{1}+\frac{\epsilon}{2}(\nabla\delta\theta_{1})^2+
	f_3^\prime 
	\delta\theta_{1}\Big)
	- f_1^\prime 
	\delta\theta_{1}
	-\frac{\epsilon}{\bar{\lambda}(t)} \dot f^\prime_1
	 \Big(\partial_t\delta\theta_{1}+\frac{\epsilon}{2}(\nabla\delta\theta_{1})^2+
	 f_3^\prime 
	 \delta\theta_{1}\Big)\delta\theta_{1}=0.
	\label{gravitydynamicsflrw}
	\end{split}
	\end{equation}
\end{widetext}

Solving this nonlinear partial differential equation for a given $\lambda(t)$, even 
for the simplest case of instantaneously switching from zero to a finite $\lambda$, where it simplifies considerably, 
and proving the convergence of the solution is a nontrivial task. 
We defer this to a future study. 
Solving similar classes of partial differential equations is performed for example by the 
Poincar\'e-Lighthill-Kuo method \cite{Nayfeh2008perturbation}.  
We can however conclude already from the level of complexity of Eq.~\eqref{gravitydynamicsflrw} 
that the renormalized metric will show 
features not present in the bare  
FLRW metric.

At late times, the perturbative expansion 
breaks down, and only a fully nonlinear treatment similar to the classical analysis of Ref.~\cite{datta2022analogue} will thereafter be applicable. Finally, a shock will be encountered, and dispersive effects, coming in a BEC from density gradients (the quantum pressure which we neglected), need in addition to be considered \cite{dattashock}.

\section{Conclusion}
The intrinsic nonlinearity of every interacting quantum field theory (classically and in the hydrodynamical  limit a fluid-dynamical system with a given equation of state) requires in general that one goes beyond the standard linearized Unruh paradigm when formulating an analogue gravity scenario.
{Specifically,} to account self-consistently for effects  which stem from the nonlinearity,  
to second (or higher) order in the {fluctuation amplitude}, one needs to take {\em backreaction} 
of the fluctuations onto the condensate into account. 
We have demonstrated   
in 
our approach to the quantum many-body problem that this engenders, when enforcing Lorentz invariance of the wave equation for the second-order fluctuations,  
{to leading order in the many-body expansion in powers of $N^{1-/2}$,}  
corrections to the Unruh paradigm metric (which is by definition seen by the first-order fluctuations). 
To higher order in $N^{-1/2}$, quantum nonlinearity, in addition, creates emergent vector fields. 
We have furthermore shown that number conservation implies 
the potential presence of a imaginary part of the frequency for excitations above {the 
\textit{\textbf{background-II}} for second-order perturbations}. 

While probing these correcting nonlinear effects beyond the standard analogue gravity Unruh paradigm 
will be experimentally challenging, they might in future experiments become testable. {For example 
by increasing, in the highly controlled way feasible in the quantum optical context, 
the contact coupling $\lambda$ via Feshbach resonances to simulate quantum many-body effects on 
a FLRW background as discussed above.} 
This, then, takes the detail to which quantum field dynamics in curved spacetime, and thus the background field concept itself, can be tested to another level of accuracy. In experimental terms, this is contingent on the 
ability to measure higher-order correlation functions than those of the (standard) density-density variety, as experimentally demonstrated, e.g., in Ref.~\cite{Schweigler}. 

\section{ACKNOWLEDGMENTS}

We thank Satadal Datta and Caio C\'esar Holanda Ribeiro for helpful discussions. 
{KP would like to acknowledge Arun Rana and Kuntal Pal for instructive conversations.} 
This work has been supported by the National Research Foundation of Korea under 
Grants No.~2017R1A2A2A05001422 and No.~2020R1A2C2008103.

\addappendix



We show here how the quantum backreaction term $\phi_{2}$ 
affects the evolution of a quantum-backreacted condensate. 
 According to the fluid dynamical description of the BEC, which we have employed here, the primary two equations governing the dynamics are  \eqref{Continuity equation} and \eqref{Bernoulli equation}. These two equations in terms of the classical fluid variables are different from the 
 continuity and  Euler equations obtained from the  
 GPE \eqref{Eq:GrossPitaevskii} due to the presence of $f_{1}$ and $f_{3}$, which 
 encapsulate the backreaction contribution from the depletion cloud. Here, our primary goal is an estimation of the size and development of these contributions in a realistic model of a BEC which accounts for backreaction.  
We will to this end consider the model studied previously in \cite{baak2022number}, where an initially noninteracting Bose gas of uniform density at rest, trapped in a hard-walled one-dimensional container  (i.e., in a one-dimensional box trap with no lateral movement permitted by the trapping geometry) of total length $\ell$ extending from $x=-\ell/2$ to $x=+\ell/2$, is subjected to an interaction quench at $t=0$. 
For $t<0$, the gas is described as solution of the GPE $\phi_{ 0}$ such that $\rho_{0}=|\phi_{ 0}|^2=$\,constant and is at rest. In the noninteracting regime ($\lambda=0$), the density of the depleted cloud $\rho_{\hat{\chi}}=0$, and consequently backreaction is also not present, i.e., $\phi_{2}=0$. For $t\geq 0$,  
 the leading-order condensate order parameter $\phi_{ 0}$ remains unchanged, and quantization of the field $\hat{\chi}$, along with the condensate-correction described by $\phi_{2}$, describes the coupled evolution of the condensate and noncondensate parts analytically  \cite{Baak:2024ajn, Ribeiro:2024vem}.

To compute the function $f_{1}(x,t)$ related to the real part of the anomalous correlator, we resort to the continuity equation satisfied by the background $\phi_{ c}$, Eq.~\eqref{backgroundcontinuity}, to obtain 
\begin{equation}
f_{1}(x,t)=\partial_{t}\rho_{\phi_{c}}+\partial_{x}J_{\phi_{c}}.
\label{f1}
\end{equation}
{We then use that for the solution of Ref.~\cite{Baak:2024ajn} we have 
$\rho_{\phi_c}=\rho_{\phi_0}+\rho_{\phi_2}+\mathcal{O}(N^{-1/2})$ and $J_{\rm c}=J_{\phi_0}+J_{\phi_2}+\mathcal{O}(N^{-1/2})$, where $J_{\phi_0}=\mbox{Im}[\phi^*_0\partial_x\phi_0]$ and 
\begin{align}
\rho_{\phi_2}&=2\mbox{Re}\ [\phi_0^*\phi_2],\label{rhozeta1}\\
J_{\phi_2}&=%
\mbox{Im}\ [\phi_0^*\partial_x\phi_2+(\partial_x\phi_0)\phi_2^*].\label{jzeta}
\end{align}
Since the analytical solution for $\phi_{ c}$ in terms of $\phi_{ 0}$ and $\phi_{2}$
is known, it is possible to compute $f_{1}(x,t)$ from \eqref{f1}, setting 
$\phi_0=\exp(-i\mu_0 t)\sqrt{\rho_0}$, where $\mu_0 = \lambda\rho_0$ is the (lowest order) chemical potential, and 
the lowest-order density $\rho_0$ is constant, which defines the bulk coherence length scale via $1/\xi_0^2 = \lambda\rho_0=c_{s0}^2$, so that all quantities are expressed in powers of $\xi_0$.}

For being self-contained, we quote here in full the form of {the second-order contributions to density 
$\rho_{\phi_{2}}$ and current $J_{\phi_{2}}$, for $t\geq0$ {(where $t$ is in units of $\xi_0^2$})},  from Ref.~\cite{baak2022number}   
\begin{widetext}
\begin{eqnarray}
\rho_{\phi_{2}} 
&=&-\frac{t^2}{\ell}-\frac{1}{4\ell}\sum_{n=1}^{\infty}\frac{(-1)^n}{\omega_n^2}\left\{2(-1)^n[1-\cos(2\omega_n t)]+\cos(2k_n x)\left[\frac{2-k_n^2}{1+k_n^2}+2\cos(2\omega_n t) 
-\frac{k_n^2+4}{k_n^2+1}\cos(\omega_{2n}t)\right]\right\}\!,\label{rhophi2} \\
J_{\phi_{2}}&=&
\frac{-2}{\ell}\sum_{n=1}^\infty\frac{(-1)^n\sin(2k_nx)}{k_n}\left[\frac{\sin(2\omega_nt)}{2\omega_n}-\frac{\sin(\omega_{2n}t)}{\omega_{2n}}\right].\label{jphi2}
\end{eqnarray}
\end{widetext}
These expressions can be obtained from the backreaction field $\phi_{2}$, which was solved for exactly in \cite{baak2022number}.  Here, we have indicated mode frequencies $\omega_n=\sqrt{k_n^2(k_n^2/4+1)}$, and wavevectors $k_n=n\pi/\ell$, which are a result of imposing Neumann boundary conditions at $x=\pm\ell/2$.

\begin{figure}[t]
\centering
\subfigure[]{\includegraphics[width=0.48\textwidth]{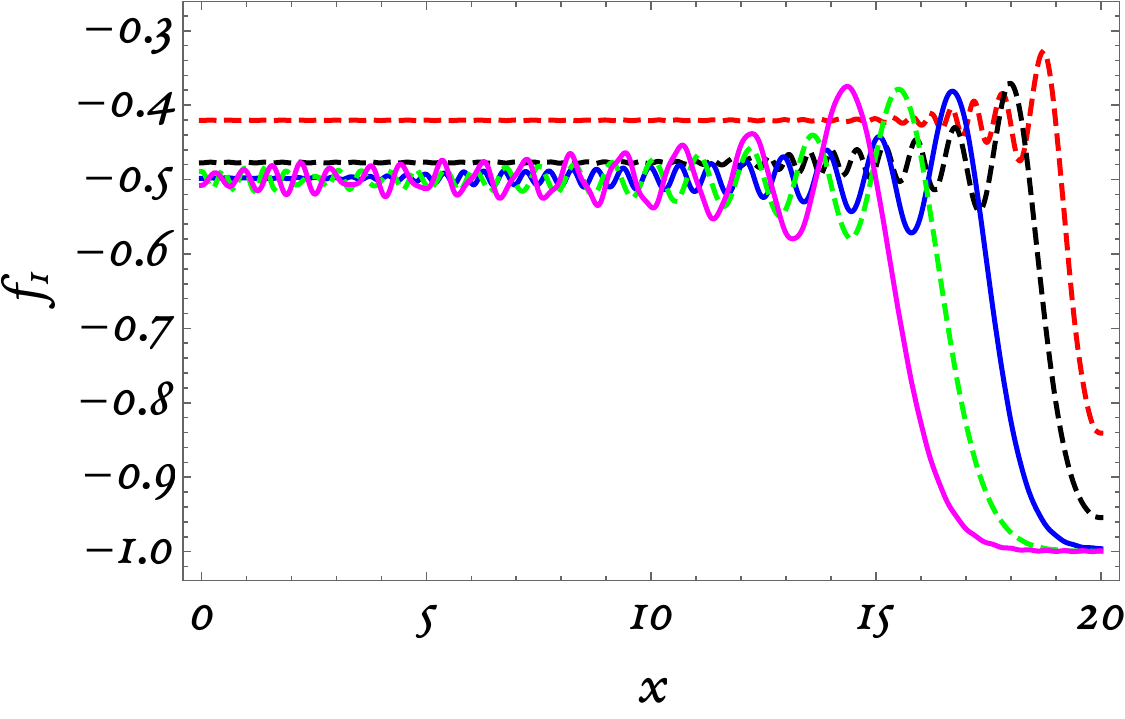}}
\subfigure[]{\includegraphics[width=0.48\textwidth]{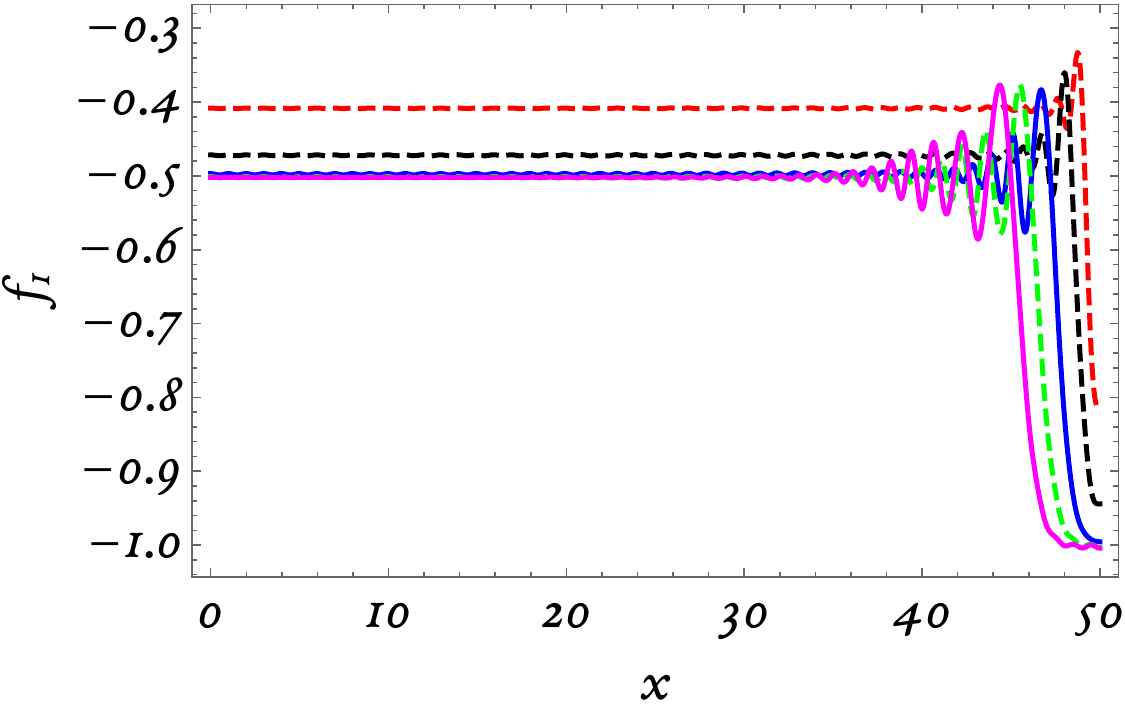}}
\caption{(a) Temporal evolution of the spatial dependence of $f_{1}$ for a condensate of total 
length $\ell=40$, extending from $x=-20$ to $x=20$. 
Red:  $t=0.5$, Black: $t=1$, Blue: $t=2$, Green: $t=3$, Magenta: $t=4$.
  (b) Evolution of  $f_{1}$ for a condensate of larger size $\ell=100$; instants of time indicated
    with identical colors as in (a). We have verified that the plots 
    do not visibly differ for mode numbers $n\gtrsim 300$; 
    {$x$ is in units of $\xi_0$ and $f_1$ is in units of $1/\xi_0^3$.} 
   \label{f1plots}} 
    \end{figure}
    
 \begin{figure}[b]
\centering
\includegraphics[width=0.48\textwidth]{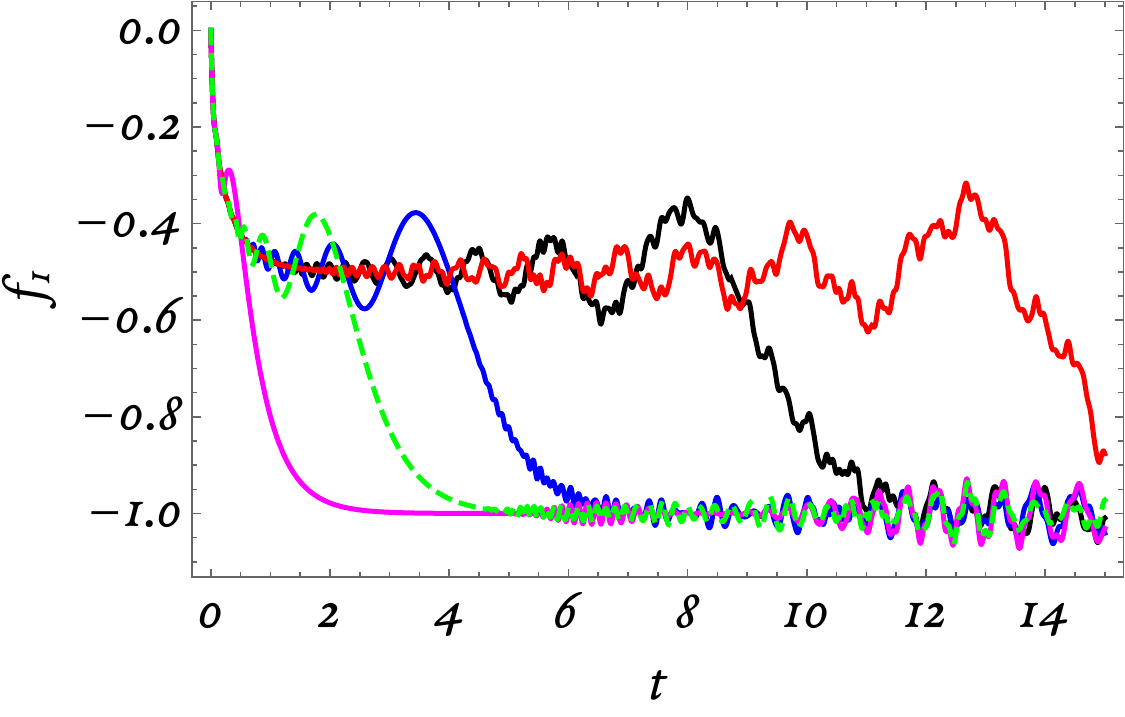}
\caption{Temporal evolution of $f_{1}$ for a condensate of total length $\ell=40$ at several locations $x$, parameters as in Fig.~\ref{f1plots} (a).
Red:  $x=5$, Black: $x=10$, Blue: $x=15$, Green: $x=17$, Magenta: $x=19$. Note that just after the interaction is turned on, there is a universal growth of particle loss from the condensate, irrespective of the position in the trap; 
    {$x$ is in units of $\xi_0$ and $f_1$ is in units of $1/\xi_0^3$.} }
    \label{f1plotsx}
    \end{figure}

We present in Figs.~\ref{f1plots} (a) and (b)  
selected plots of the variation of $f_{1}(x,t)$ with $x$ at various instants $t\geq0$.
We note that for all profiles the function $f_{1}<0$, which 
confirms that there is an outgoing particle flux leaving the condensate (particle loss by the condensate). 
After the system is quenched from a noninteracting to an interacting regime, $f_{1}$ becomes nonzero, following a pattern similar to that of $\rho_{\hat{\chi}}$ \cite{baak2022number}, and such that the growth in the bulk of the system is not sensitive to the boundary region. 

We also plot in Fig.~\ref{f1plotsx} the evolution of $f_{1}$ with time for different locations inside the trap. We infer that the growth of $f_1$ at different $x$, just after the interaction is turned on, follows the same universal pattern. It is dominated by a quadratic growth, which is the contribution of the zero mode 
contained in \eqref{rhophi2}. This growth is universal irrespective of the location inside the trap. 
Furthermore, the evolution is drastically different depending on the location in the trap. 
Near the center of the gas, it is oscillatory. However, close to the hard wall, the contribution to the backreaction quickly saturates with time. These observations are consistent with the observations made in Ref.~\cite{Ribeiro:2024vem}, that as the interaction is suddenly turned on, the energy exchange of the condensate with the quasiparticle excitation cloud initially occurs very fast, and then 
saturates to an oscillatory behavior, within a time scale depending on the position in the trap.


\bibliography{amr10.bib}

\end{document}